\documentstyle{article}
\title{
{\small
\begin{flushright}
UUITP 8/1994 \\
16pp, Latex\\
$mp_arc/9404161$\\
\copyright \\
\end{flushright}}
\bigskip \bigskip \bigskip \bigskip \bigskip \bigskip
 Uniform Asymptotic Solutions of a System of two Schr\"odinger
Equations with Potential-Curve-Crossing Point.}
\author{ Irina Jakushina\thanks{On leave from: St.Petersburg Academy
of Air-Space Technology,198000 St.Petersburg,
Gertsena,67, Russia} \\
               Department of Theoretical Physics \\
        University of Uppsala \\
        Box 803
        S-751 08 Uppsala, Sweden\\}

\vspace{2cm}

\date{2 April}

\begin{document}

\renewcommand{\baselinestretch}{1}
\large
\maketitle

\vspace{3cm}
\begin{abstract}

A formal uniform asymptotic solution of the system of differential
equations
$ h^{2}\frac{d^{2}U_{1}}{dz^{2}}+\Phi_{1} U_{1}=\alpha U_{2} $ , $
h^{2}\frac{d^{2}U_{2}}
{dz^{2}}+\Phi_{2} U_{2}=\alpha U_{1}$ , for $ z\in D$ and for h real,
large is
obtained, when D contains curve-crossing point.  Asymptotic
approximations for
the solutions are constructed in terms of parabolic cylinder
functions. Analytical
properties of the expansion's coefficients are investigated.The case
of potantial barier is also considered.

\end{abstract}

\newpage

\section{Introduction}
\label{sec:intro}
We  consider the system of differential equations
\begin{eqnarray}
h^{2}\frac{d^{2}U_{1}}{dz^{2}}+\Phi_{1}(z)U_{1}=\alpha(z,\delta)
U_{2}\nonumber\\
h^{2}\frac{d^{2}U_{2}}{dz^{2}}+\Phi_{2}(z)U_{2}=\alpha(z,\delta)U_{1}
\end{eqnarray}

Here $\Phi_{1},\Phi_{2},\alpha$ are functions of the complex variable
$z$ in the domain $ D$ and of the parameter $ \delta $ in the complex
domain G $(0\in G)$.
We suppose, that $\alpha(z,0)\equiv 0$.
The main subject of this paper is the construction of the uniform
asymptotic expansion
of the solution $U_{1}$ for the parameter h small and positive in the
region $\Omega=D\times G$, which
contains a crossing point $z=x_{0}$ (a point, where
$\Phi_{1}(x_{0})=\Phi_{2}(x_{0})$).The last part of the paper only is
devoted to potantial barier problem in.

\vspace{.1in}

    The problem of energy level crossing is of considerable practical
importance. It occurs in
different branches of physics ( see e.g.\cite {aq:lan}\cite {he:iss})
, but it doesn't have still a  general analytical solution.
First it was investigated by phase integral methods by Stueckelberg
\cite {st:tuc} and from time
 dependent point of view by Landau \cite {la:dau} . The theory has
also been developed by analysis
in the momentum representation by Bykovskii{\em et. al.} \cite
{bu:kov} . Later it was treated by  Crothers {\em et. al.}
\cite {cr:rot}\cite {co:ver}
 but only the model problem with linear potentials and constant
coupling was considered.

\vspace{.1in}

     In this paper we use the comparison equation technique
\cite{an:yan} \cite {ol:ver}
  and we establish uniform
asymptotic approximations for the solutions in terms of parabolic
cylinder functions.
One important property of the 4-th order equation for the function
$U_{1}$ is that
the coefficients of the equation as well as turning points depend on
h. At the case,
 when $\delta=0$ ($\alpha\equiv 0$), the turning points don't
coalesce exactly. The distance
between them is proportional to h.
In \cite {ol:ver} it is shown how to construct the first member of
the expantion for the solutions
of the second order differential equation with two close turning
points and coefficients
depending on h. However, the case of a higher order equation and
arbitrary order
 approximations has not been pursued. For the comprehention of the
problem considered the work
of Fedoruk M.V.\cite {fe:ruk}  is important.

\vspace{.1in}

We don't consider the asymptotic nature of the formal solution in the
present paper,

but the analytical properties of the expantion's coefficients are
investigated.

\vspace{.3in}

\section{WKB-solutions}
\label{sec:nonlinosc}

\vspace{.1in}

{}From(1)we get the fourth-order
differential equation for the function $  U_{1}$

\begin{eqnarray}
U_{1}^{''''}+\frac{1}{h^{2}}
U_{1}^{''}(\Phi_{1}+\Phi_{2})+\frac{1}{h^{4}}U_{1}(\Phi_{1}
\Phi_{2}-\alpha^{2})+\nonumber\\
2\alpha(\frac{1}{\alpha})^{'}U_{1}^{'''}+2\alpha(\frac{\Phi{1}}{\alpha
})^{'}\frac{1}
{h^{2}}U_{1}^{'}+\nonumber\\
\alpha(\frac{1}{\alpha})^{''}U_{1}^{''}+
\frac{1}{h^{2}}\alpha(\frac{\Phi_{1}}{\alpha})^{''}U_{1}=0
\end{eqnarray}

The solutions of equation (2) when D doesn't contain any turning
points are
well-known \cite {fe:ruk}

\begin{eqnarray}
y_{1,2}(z)=\frac{\sqrt{\sqrt{D}-\Psi}}{\sqrt[4]{P_{10}^{2}D}}\exp(\pm\
frac{1}{h}
\int^{z}p_{10}(t)dt)\nonumber\\
y_{3,4}(z)=\frac{\sqrt{\sqrt{D}-\Psi}}{\sqrt[4]{P_{10}^{2}D}}\exp(\pm\
frac{1}{h}
\int^{z}p_{30}(t)dt)
\end{eqnarray}

In (3,4) we have introduced the notations $p_{i0}(z),(i=1,2,3,4)$ for
the roots of
the equation
\begin{eqnarray}
 l_{0}(z,p,\lambda)\equiv
p^{4}+(\Phi_{1}+\Phi_{2})p^{2}+(\Phi_{1}\Phi_{2}-\alpha^{2})=0
\end{eqnarray}

This roots have the form

\begin{eqnarray}
p_{10,20}=\pm\sqrt{\Phi(z)+\sqrt{\Psi^{2}(z)+\alpha^{2}}},\nonumber\\
p_{30,40}=\pm\sqrt{\Phi(z)-\sqrt{\Psi^{2}(z)+\alpha^{2}}}
\end{eqnarray}
where
\begin{eqnarray}
\Phi(z)=-\frac{1}{2}[\Phi_{1}(z)+\Phi_{2}(z)],\nonumber\\
\Psi(z)=\frac{1}{2}[\Phi_{1}(z)-\Phi_{2}(z)]\nonumber\\
D=\Psi^{2}+\alpha^{2}
\end{eqnarray}
\vspace{.3in}

\section{Equation with fixed turning points}
\label{sec:pitheory}

On extracting the largest members in terms of h from the equation (2)
we get the equation

\begin{eqnarray}
U_{1}^{''''}+\frac{1}{h^{2}}
U_{1}^{''}(\phi_{1}+\phi_{2})+\frac{1}{h^{4}}U_{1}(\Phi_{1}
\Phi_{2}-\alpha^{2})=0,
\end{eqnarray}
the coefficients of which are independent from the parameter h. For
the functions $\Phi_{1},
\Phi_{2},\alpha$ we make the following assumptions:1.$\Phi_{1},
\Phi_{2}$ are analitic functions for all $z\in D$, $\alpha$ is an
analytic function for all
$(z,\delta)\in \Omega $     2.At the point $x_{0}\in D$ we have
$\Phi_{1}(x_{0})=
\Phi_{2}(x_{0}) \neq 0 $   3. $\alpha(x,0)\equiv 0$.

The symbol of the equation (2) is of the form (4) and the roots
$p_{i0} (i=1,...4) $ of the equation
$ l_{0}(z,p,\delta)=0$ are given by (5).

\vspace{.1in}

 When the
function $\alpha(z)$ is small, the equation $
(\Phi_{1}-\Phi_{2})^{2}+4\alpha^{2}=0$
has two roots $z_{10}$ and $z_{20}$ $ (z_{10}=\overline {z_{20}})$ in
the neighborhood of
the point $x_{0}$.Then $ p_{10}(z_{i0})=p_{30}(z_{i0})$ and $
p_{20}(z_{i0})=p_{40}(z_{i0})$
and $z_{i0}$ are turning points of the equation(7).
 When $ \delta=0$ the turning points $ z_{10}$ and $z_{20}$ coinside
and are equal to $ x_{0}$.
We further assume, that the region D doesn't contain any other
turning points of the equation(7).

\vspace{.2in}

We can also represente $l_{0}$ in the following way:

\begin{eqnarray}
l_{0}(p,h,\delta)=(p^{2}+a_{3}p+a_{2})(p^{2}+a_{1}p+a_{0}),
\end{eqnarray}

where

\begin{eqnarray}
a_{3}=-(p_{10}+p_{30})=-(-(\Phi_{1}+\Phi_{2})-2\sqrt{\Phi{1}\Phi{2}-\a
lpha^{2}})^{1/2},\\
a_{2}=p_{10}p_{30}=2(-\Phi_{1}\Phi_{2}-\alpha^{2})^{1/2},\\a_{1}=-a_{3
},\\a_{0}=a_{2}
\end{eqnarray}
\newtheorem{func}{Lemma}
\begin{func}

The functions $a_{i} (i=1,...4)$ are analytic for all $(z,\delta)\in
\Omega$.
\end{func}

\vspace{.1in}

Since we don`t have other turning points
$\Phi_{1}\Phi_{2}-\alpha^{2}\neq 0$.It«s easy also to
see that
$-(-(\Phi_{1}+\Phi_{2})-2\sqrt{\Phi_{1}\Phi_{2}-\alpha^{2}})^{1/2}\neq
 0$
 for all $z\in D$. Because of $\alpha^{2}>0$  $\Phi_{1}\Phi_{2}>0$,
we have
$\Phi_{1}\Phi_{2}+\alpha^{2}\neq 0$ for all $(z,\delta)\in \Omega $.
We observe, that $ a_{1}$ and $a_{2}$
 are analytic functions of z, like square roots of analytic
functions, not equal
to zero.
Both first and second brackets in (8) have multiple roots when $
z=z_{10}$ and $ z=z_{20}$
and, concequently, equation (7) has two simple turning points at $
z=z_{10}$ and at
$ z=z_{20}$.

\vspace{.1in}

We wish to obtain an asymptotic representation for any solution $
U_{1}$ of (7 )
which is uniform in D ( including the points $ z_{1} $, $ z_{2}$).

We seek the asymptotic expantions  of two linearly independent
solutions of (7 ),
corresponding to the first bracket in (8) of the form

\begin{eqnarray}
U_{i}(z,\delta,h)= \exp(\frac {1}{2h}
\int^{z}(p_{10}+p_{30})dt)\times \nonumber \\
(A E_{i}(\frac{\tau_{1}}{h},\frac{1}{\sqrt{h}}\xi_{1})+
\sqrt{h}B E_{i}^{'}(\frac{\tau_{1}}{h},\frac{1}{\sqrt{h}}\xi_{1})),
i=1,2
\end{eqnarray}

where $E_{1,2}$ are two linearly independent  solutions of the Weber
equation. The choice of these
functions depends on the sign of the functions $\Phi_{1}$, $\Phi_{2}$
in D.
Next we assume, that A,B,$\tau$, $\xi$ have the representations

\begin{eqnarray}
A(z,\delta,h)=\sum_{i=1}^{\infty}a_{i}(z,\delta)h^{i}
\\ B(z,\delta,h)=\sum_{i=1}^{\infty}b_{i}(z,\delta)h^{i}\\
 \tau(\delta,h)=\sum_{i=1}^{\infty}\tau_{i}(\delta)h^{i}\\
\xi(z,\delta,h)=\xi(z,\delta)
\end{eqnarray}

On employing the anzats (13) in (7) we get the following expresions
for $ \tau_{0}$ and $\xi$ \cite {ja:kus}

\begin{eqnarray}
\tau_{0}(\lambda)=\frac{-1}{2\pi}\int_{z_{1}(\delta)}^{z_{2}(\delta)}\
sqrt{F}dt\\
\int_{2i\sqrt{\tau_{0}(\delta)}}^{\xi(z,\delta)}\sqrt{-\xi^{2}/4-\tau_
{0}(\delta)}d\xi=
\frac{1}{2}\int_{z_{2}(\delta)}^{z}\sqrt{F}dt,
\end{eqnarray}

where $F=\frac{1}{4}(p_{10}-p_{30})^{2}$

We choose the branches of the roots here in the following way:
$\sqrt{F}\geq 0$ for$F\geq 0$ .

\begin{func}

Equation (19) defines function $\xi(z,\delta)$ for all $z\in D$,
$\delta\in G$
with the following properties:
1. $\xi$ is analytic in $ D\times G$
2. $\xi(\pm2i\sqrt{\tau_{0}})=z_{1,2}$
3. $\xi^{'}(z,\delta)\neq 0$ for all $ z\in D$,$\delta\in G$.

\end{func}
\vspace{.1in}

The proof of this lemma based on Hartogs theorem(see \cite {ja:kus})

In the next order of approximation we get the expresions for $a_{0}$
and $b_{0}$ \cite {ja:kus}:

\begin{eqnarray}
a_{0}(z)=k\exp(\int_{z_{1}}^{z}\Psi_{1}(t)dt)\cosh(\int_{z_{1}}^{z}(\P
si_{2}(t)-
\frac{i}{2}\tau_{1}\frac{\xi^{'}}{\sqrt{\xi^{2}/4+\tau_{0}}})dt)\\
b_{0}(z)=\frac{k}{\sqrt{-\xi^{2}/4-\tau_{0}}}
\exp(\int_{z_{1}}^{z}\Psi_{1}(t)dt)\sinh(\int_{z_{1}}^{z}(\Psi_{2}(t)-
\nonumber\\
\frac{i}{2}\tau_{1}\frac{\xi^{'}}{\sqrt{\xi^{2}/4+\tau_{0}}}dt
\end{eqnarray}
where
\begin{eqnarray}
\Psi_{1}(z)=-\frac{\xi^{''}}{\xi^{'}}+
a^{'}\sum_{k=2,4}\frac{q_{k}}{q_{k}^{2}-F}+
F^{'}\sum_{k=2,4}\frac{1}{q_{k}^{2}-F}\\
\Psi_{2}(z)=\frac{1}{2\sqrt{F}}(-a^{'}+a^{'}\sum_{k=2,4}\frac{F}{q_{k}
^{2}-F}+
F^{'}\sum_{k=2,4}\frac{q_{k}}{q_{k}^{2}-F})\\
a=\frac{1}{2}(p_{10}+p_{30}),
q_{k}=p_{k}-a
\end{eqnarray}

The parameter $\tau_{1}$ we can find from the condition, that
function $b_{o}(z)$
is analytic at $z=z_{2}$
\begin{equation}
\tau_{1}=-\frac{1}{\pi}\int_{z_{1}}^{z_{2}}\Psi_{2}(t)dt
\end{equation}

As we have seen, the roots $p_{20}$,$p_{40}$ have branch points at
the turning
points $z_{10}$, $z_{20}$ . But the functions $\Psi_{1}$
and$\Psi_{2}$ depend only
on expressions $p_{20}+p_{40}$ and $p_{20}p_{20}$, which are analytic
functions
on$ (z,\delta)$. This allows us to prove the following lemma.

\vspace{.2in}
\begin{func}
 The coefficients $a_{0}$, $b_{0}$, defined by (20),(21) are analytic
for all
$(z,\delta)\in D\times G$.
\end{func}

\vspace{.2in}

The proof of the lemma in the case, when the second bracket in (8)
doesn't have multiple
zeros in D is done in \cite {ja:kus}  and it's easy to employ this
proof to this case also.

\vspace{.1in}

The other two linearly independent solutions of equation (7) we can
get by changing the numbers
šfor the roots 1,3 to2,4 in formulas (18)-(24).

\vspace{.2in}

\section{General equation}
\label{sec:dhyrml}

Now we shall consider the differential equation (2 ), the
coefficients and
 turning points of which
depend on the small parameter h .The symbol of the equation (2) has
the form
\begin{eqnarray}
 l(z,p,\lambda)=p^{4}+(\Phi_{1}+\Phi_{2})p^{2}+(\Phi_{1}\Phi_{2}-\alph
a^{2})+
2\alpha h(\frac{1}{\alpha})^{'}p^{3}+2\alpha
h(\frac{\Phi{1}}{\alpha})^{'}\frac{1}
{h^{2}}p+\nonumber\\
\alpha h^{2}(\frac{1}{\alpha})^{''}p^{2}+
\frac{1}{h^{2}}\alpha h^{2}(\frac{\Phi_{1}}{\alpha})^{''}
\end{eqnarray}
The roots of the characteristic equation $l(z,p,\delta)=0$ also
depend on the parameter h:
$p_{i}=p_{i}(z,h)(i=1,2,3,4)$
as well as turning points
$z_{i,k}(h),i=1,2;k=1,2$
Here $z_{i,1}(h)(i=1,2)$ are the roots of the equation
$p_{1}(z,h)=p_{3}(z,h)$ and
 $z_{i,2}(h)$ are the roots of the
equation $p_{2}(z,h)=p_{4}(z,h)$.
In the previous part we used only the roots $p_{i}$ of the
caracteristic equation
 for constructing the asymptotical solution of the equation (7 ). Now
 we can repeate all calculations and write the asymptotic solution of
(2 ) in terms of $p_{i}(z,h)$.
But it's impossible to use this solution, because  it's impossible to
find exactly
the roots $p_{i}(z,h)$ of 4-th order equation $l(z,p,\delta)=0$. We
also cannot
expand the roots $p_{i}(z,h)$, because they have branch points at
turning points
 of the equation (2 ). But  the expressions $p_{1}(z,h)+p_{3}(z,h)$,
 $p_{1}(z,h)p_{3}(z,h)$, $p_{2}(z,h)+p_{4}(z,h)$ and
 $p_{2}(z,h)p_{4}(z,h)$ are analytic functions for all $(z,\delta)\in
\Omega $,$ h<\varepsilon$
like it was for the roots of the equation (7 ).It is shown in the
follšwing lemma

\vspace{.2in}
\begin{func}
Consider the expression
\begin{eqnarray}
l(p)=(p^{2}+a_{3}p+a_{2})(p_{2}+a_{1}p+a_{0})+c_{3}p_{3}+c_{2}p_{2}+c_
{1}p+c_{0}
\end{eqnarray}
where $a_{i},c_{i},i=1,...4$ are analytic functions for all
$(z;\delta)\in D\times G$ ,$c_{i}=O(h)$, $c_{i}$ are analytic on h
for $ h<\varepsilon$.
Let $p_{i}\neq p_{k}, i=1,2, k=3,4$ for all $(z,\delta)\in D \times
G$.
Then functions $\alpha_{i}, i=1,2,3,4$ exist and are analytic
for all $(z,\delta)\in D\times G,h< \varepsilon$ such,that
\begin{eqnarray}
l(p)=(p^{2}+(a_{3}+\alpha_{3})p+(a_{2}+\alpha_{2}))
(p_{2}+(a_{1}+\alpha_{1}p+(a_{0}+\alpha_{0}))
\end{eqnarray}

Let the vektor \hbox{\boldmath$\alpha$}  be $\hbox{\boldmath$\alpha$}
=(\alpha_{3},\alpha_{2},\alpha_{1},
\alpha_{0})^{t}$. Then
for the coefficient$\hbox{\boldmath$\alpha$} _{0}$ in the expansion
$\hbox{\boldmath$\alpha$} =\hbox{\boldmath$\alpha$ }_{0}h+.....$
we have
\begin{equation}
\hbox{\boldmath$\alpha$} _{0}=M^{-1}{\bf c}
\end{equation}
where M is the matrix
\begin{equation}
M=\left(\matrix{
1&1&0&0\cr
a_{1}&a_{3}&1&1\cr
a_{0}&a_{2}&a_{1}&a_{3}\cr
0&0&a_{0}&a_{2}\cr
}\right)
\end{equation}
\end{func}

\vspace{.2in}

{\em Proof of the lemma.} Let \hbox{\boldmath$\beta$}  be the
vector$( 0, \alpha_{3}\alpha_{1},
\alpha_{3}\alpha_{0}+ \alpha_{2}\alpha_{1},
\alpha_{2}\alpha_{0})^{t}$. From (27)
and (28)we get the equation
\begin{equation}
M \hbox{\boldmath$\alpha$} +\hbox{\boldmath$\beta$} =\bf c
\end{equation}
The determinant of the matrix M is
\begin{eqnarray}
detM=
(a_{3}-a_{2})(a_{2}a_{1}-a_{3}a_{0})-a_{2}-a_{0}^{2}=\nonumber\\
-(p_{3}-p_{1})(p_{3}-p_{2})(p_{4}-p_{1})(p_{4}-p_{2})
\end{eqnarray}

We observe, that the $detM=0$ only when $p_{i}=p_{k},i=1,2,k=3,4$.
 Because of the conditions of the lemma $detM\neq0$ and from (31) we
obtain the equation
\begin{equation}
\hbox{\boldmath$\alpha$} =M^{-1}{\bf c}-M^{-1}\hbox{\boldmath$\beta$}
\end{equation}
As we have suppose, ${\bf c}=h \tilde{\bf c}$.Then
$\hbox{\boldmath$\alpha$} =h\tilde{\hbox{\boldmath$\alpha$}}$ and
 $\hbox{\boldmath$\beta$} =h^{2}\tilde{\hbox{\boldmath$\beta$}}$
For sufficiently small h $(h<\varepsilon)$ we can solve the equation
\begin{equation}
\tilde{\hbox{\boldmath$\alpha$}}=M^{-1}{\bf
\tilde{c}}-hM^{-1}\tilde{\hbox{\boldmath$\beta$}}
\end{equation}
by successive approximations. The solution
$\tilde{\hbox{\boldmath$\alpha$}}$ is analytic function
for all$(z,\delta)\in D\times G
                                $
                                 and $h<\varepsilon$, because  the
function
$\bf c$ is analytic.The first order of approximation in h for
\hbox{\boldmath$\alpha$}  we can find
 from (29).
\vspace{.2in}

In the case of curve-crossing we have the symbol of the equation:
\begin{eqnarray}
l(z,p,\delta)=(p^{2}-(p_{10}+p_{30})p+p_{10}p_{30})(p^{2}-(p_{20}+p_{4
0})p+\nonumber\\
p_{20}p_{40})+ c_{3}p^{3}+ c_{1}p+....=\nonumber\\
(p^{2}-(p_{1}+p_{3})p+p_{1}p_{3})(p^{2}-(p_{2}+p_{4})p+
p_{2}p_{4})
\end{eqnarray}
where $ c_{3}= 2\alpha(\frac{1}{\alpha})^{'}
,c_{1}=2\alpha(\frac{\Phi{1}}{\alpha})^{'}$
Emploiyng the results of Lemma 4 we see that $p_{1}+p_{3},
p_{1}p_{3}, p_{2}+p_{4} and
 p_{2}p_{4}$ are analytic functions on $(z,\delta)$ and h. We can
also find two first members
in the series on h:
\begin{eqnarray}
p_{1}+p_{3}= p_{10}+p_{30}- \frac{1}{2}c_{3}+....\nonumber\\
p_{1}p_{3}=p_{10}p_{30}+\frac{a_{2}}{2a_{3}}c_{3}-\frac{1}{2a_{3}}c_{1
}+...\nonumber\\
p_{2}+p_{4}= p_{20}+p_{40}- \frac{1}{2}c_{3}+....\nonumber\\
p_{2}p_{4}=p_{20}p_{40}+\frac{a_{0}}{2a_{1}}c_{3}-\frac{1}{2a_{1}}c_{1
}+...
\end{eqnarray}
The function$
(p_{1}-p_{3})^{2}$ and $
(p_{2}-p_{4})^{2}$ are also analytic on h and z, and we can find two
first
members of the expansion
\begin {eqnarray}
(p_{1}-p_{3})^{2}= f_{1}(t,h,\delta) +O(h^{2}),\\
(p_{2}-p_{4})^{2} =f_{2}(t,h,\delta)+O(h^{2}),
\end{eqnarray}
where
\begin{eqnarray}
f_{1}(t,h,\delta)=(p_{10}-p_{30})^{2}-\frac{2h}{p_{10}+p_{30}}(\alpha(
\frac{1}{\alpha})^{'}
(p_{10}^{2}+p_{30}^{2})+2\alpha(\frac{\Phi_{1}}{\alpha})^{'}) \\
f_{2}(t,h,\delta)=(p_{20}-p_{40})^{2}-\frac{2h}{p_{20}+p_{40}}(\alpha(
\frac{1}{\alpha})^{'}
(p_{20}^{2}+p_{40}^{2})+2\alpha(\frac{\Phi_{1}}{\alpha})^{'})
\end{eqnarray}

Because the right hand side of (39),(40) has  zeros at the turning
points
of the equation (2 ) , $ p_{1}-p_{3} $ and $ p_{2}-p_{4} $ aren't
analytic on h and z. On substituting
$p_{1}+p_{3},(p_{1}-p_{3})^{2},p_{2}+p_{4},(p_{2}-p_{4})^{2}$,
defined by (36-38)
 instad of
$p_{10}+p_{30},(p_{10}-p_{30})^{2},p_{20}+p_{40},(p_{20}-p_{40})^{2}$
we get the asymptotic solution of the equation (2 ).

\begin{eqnarray}
U_{1,2}(z)=\exp(\frac{1}{2h}\int^{z}(p_{10}+p_{30}-h\alpha(\frac{1}{\a
lpha})^{'})dt
(a_{0}U(\pm\frac{i\tau_{1}}{h},e^{\pm\frac{i\pi}{4}}\frac{1}{\sqrt{h}}
\xi_{1})+\nonumber\\
\sqrt{h}b_{0}
U^{'}(\pm
i\frac{\tau_{1}}{h},e^{\pm\frac{i\pi}{4}}\frac{1}{\sqrt{h}}\xi_{1})(1+
O(h))\\
U_{3,4}(z)=\exp(\frac{1}{2h}\int^{z}(p_{20}+p_{40}-h\alpha(\frac{1}{\a
lpha})^{'})dt
(a_{0}U(\pm\frac{i\tau_{2}}{h},e^{\pm\frac{i\pi}{4}}\frac{1}{\sqrt{h}}
\xi_{2})+\nonumber\\
\sqrt{h}b_{0}
U^{'}(\pm
i\frac{\tau_{2}}{h},e^{\pm\frac{i\pi}{4}}\frac{1}{\sqrt{h}}\xi_{2})(1+
O(h))
\end{eqnarray}
where $\xi_{i}(z,\delta,h)$ is defined by
\begin{eqnarray}
\int_{2i\sqrt{\tau_{i0}(\delta,h)}}^{\xi_{i}(z,\delta,h)}\sqrt{-\xi^{2
}/4-\tau_{i0}(\delta,h)}d\xi=
\int_{z_{i2}(\delta,h)}^{z}\sqrt{f_{i}(t,h,\delta)}dt
\end{eqnarray}
The turning points $z_{i1}(h)$ and$z_{i2}(h)$ are the roots of the
equation $f_{i}(t,h,\delta)=0$
The parameter $\tau_{0}(h,\delta)$ is defined by
\begin{equation}
\tau_{i0}(h,\delta)=-\frac{1}{2\pi}\int_{z_{i1}(\delta,h)}^{z_{i2}(\de
lta,h)}\sqrt{f_{i}(t,h,\delta)}dt
\end{equation}
The coefficients $a_{0}$ and $b_{0}$are given by (20)-(25) with
$\tau_{i0}(h,\delta)$ and
$\xi_{i}(z,\delta,h)$, defined by (43)-(44).

\section{Case $\alpha\equiv0$}
\label{sec:casee}

We now wish to specialize the asymptotic formulas(41)-(42)for the
case $\alpha \equiv 0$.
We calculate the integrals (44),(25) by residues and get
\begin{eqnarray}
\tau_{0}=-\frac{i\Phi_{1}^{'}(0)h}
{\Phi_{1}^{'}(0)-\Phi_{2}^{'}(0)}+O(h^{2})\\
\tau_{1}=\frac{i}{2}\frac{\Phi_{1}^{'}(0)+\Phi_{2}^{'}(0)}{\Phi_{1}^{'
}(0)-\Phi_{2}^{'}(0)}+O(h)
\end{eqnarray}
Finally, for the parameter $\tau$ we have
\begin{eqnarray}
\tau=\tau_{0}(h)+h\tau_{1}(h)+...=-\frac{i}{2}h+0(h^{2})
\end{eqnarray}
At the expressions(41),(42) for $U_{1,2}$ we have two Weber functions
$ U(\pm\frac{1}{2},\exp(\mp\frac{i\pi}{4})\frac{\xi_{1}}{\sqrt{h}})
$,
which are linearly dependent.
Consequantly, the solutions $U_{1}$ and $U_{2}$ are linearly
dependent.
Consider the solution $U_{1}$.Then by employing
$ U(\pm\frac{1}{2},\exp(\mp\frac{i\pi}{4})\frac{\xi_{1}}{\sqrt{h}})
=\exp(-\frac{i\xi_{1}^{2}}{4h})$ in (41) we obtain for $\xi_{1}\geq
0$
\begin{equation}
U_{1}\sim \frac{1}{\sqrt{p_{10}}}
\exp(\frac{1}{h}\int^{z}p_{10}(t)dt)
\end{equation}
and for $ \xi_{1}\leq 0 $
\begin{equation}
U_{1}\sim \frac{1}{\sqrt{p_{30}}}
\exp(\frac{1}{h}\int^{z}p_{30}(t)dt)
\end{equation}
Finally, the solution $U_{1}$ for all real z has the form
\begin{equation}
U_{1}\sim \frac{1}{\sqrt[4]{-\Phi_{1}}}
\exp(\frac{1}{h}\int^{z}\sqrt{-\Phi_{1}(t)}dt)
\end{equation}
In the same way we get for the solution $U_{3}$
\begin{equation}
U_{3}\sim \frac{1}{\sqrt[4]{-\Phi_{1}}}
\exp(-\frac{1}{h}\int^{z}\sqrt{-\Phi_{1}(t)}dt)
\end{equation}

\vspace{.3in}

\section{Potential barrier in one channel.}
\label{sec:floq}

We wish to construct the asymptotic expantions for the equation (2).
We suppose
that $ \Phi_{i}(z)=\lambda-V_{i}(z)$ ,where$\lambda$ is a
parameter,$\lambda\in G$.For The
functions $  V_{1}(z)$,$ V_{2}(z)$,$ \alpha(z)$ we make the following
assumptions
1. $ V_{1}(z)$,$ V_{2}(z)$,$ \alpha(z)$ are analytic functions in the
neighbourhood
D of the real axis.
2. Equation $ V_{1}(z)=V_{2}(z)$ doesn«t have any solutions in D (
there is no
 crossing points in D). We shall also suppose that $ V_{1}(z)   >
V_{2}(z) $ for
real z.
3. $ V_{1}(z)$ has one maximum for real z at $ z=0 $

The main symbol of the equation (2) has the form

\begin{equation}
  l_{0}(p,h,\lambda)=(p^{2}-p_{1}^{2})(p^{2}-p_{3}^{2})
\end{equation}

The turning points of the equation (2) are the points where $
l_{0}(p,h,\lambda)$
has multiple roots. Because of the assumptions 2,3 $
\Phi_{1}+\Phi_{2}>0 $
and $ p_{3}(z)\neq 0  $  in $ D$. We observe that the condition
$p_{1}=p_{2}=0  $ is
sutisfied when

\begin{eqnarray}
\Phi_{1}\Phi_{2}=\alpha^{2}.
\end{eqnarray}

The roots of the equation (7)$ x_{1}$ and$ x_{2}		$ are
tuning points of (2).
When $ \alpha=0$ turning points are zeros of the equation$
\Phi_{1}(z)=0$.
when $ |\alpha |\ll |\Phi_{2}|$ turning points are close to zeros of
$\Phi_{2}$.
For some $\lambda=\lambda_{0}$ turning points coinside and are equal
to zero.

Since asymptotic solutions $y_{3,4}(z)$ have analytic coefficients in
the
neighbourhood of turning points, this solutions are valid in D.
We wish to obtain the rest two solutions of (2)
which are uniform in D.We seek this solutions of the form

\begin{eqnarray}
y_{1,2}(z,\lambda)=A_{1,2}(z,\lambda)U(\pm\frac{i\tau_{1,2}}{h},
\exp(\pm\frac{i\pi}{4})\frac{1}{\sqrt{h}}\xi(z,\lambda))+\nonumber\\
\sqrt{h}B_{1,2}(z,\lambda)U^{'}(\pm\frac{i\tau_{1,2}}{h},
\exp(\pm\frac{i\pi}{4}\frac{1}{\sqrt{h}}\xi(z,\lambda))
\end{eqnarray}

Here $U(a,z)$ is the solution of the Weber equation
$U^{''}-(\frac{1}{4}z^{2}+a)U=0 $
and the functions A,B and $\xi$ are to be determined.Next we assume
that $A(z,\lambda,h)
$,$B(z,\lambda,h)$,$\tau(\lambda,h)$ have the representations

\begin{eqnarray}
A(z,\lambda,h)=\sum_{i=1}^{\infty}a_{i}(z,\lambda)h^{i}\\
B(z,\lambda,h)=\sum_{i=1}^
{\infty}b_{i}(z,\lambda)h^{i}\\
\tau(\lambda,h)=\sum_{i=1}^{\infty}\tau_{i}(\lambda)h^{i}
\end{eqnarray}

We employ equations (54)-(57) in (2). On equating to zero the
coefficients of
$   h^{k}U$ and $ h^{ k+1/2 } U^{'}$ we obtain

\begin{eqnarray}
\tau_{0}(\lambda)=\frac{1}{\pi}\int_{x_{1}(\lambda)}^{x_{2}(\lambda)}p
_{10}(t)dt\\
\int_{2\sqrt{\tau_{0}(\lambda)}}^{\xi(x,\lambda)}\sqrt{\xi^{2}/4-\tau_
{0}(\lambda)}d\xi=
-\int_{x_{2}(\lambda^)}^{x}p_{10}(\lambda,t)dt
\end{eqnarray}
\begin{eqnarray}
a_{10}=a_{20}=\frac{\sqrt[1/4]{\xi^{2}/4-\tau_{0}}\sqrt{\sqrt{\Psi^{2}
+\alpha^{2}}+\Psi}}{\sqrt[4]{p_{10}^{2}
(\Psi^{2}+\alpha^{2})}},\\
b_{10}=b_{20}=0,\\
\tau_{10}=\tau_{20}=0
\end{eqnarray}

That the expantion coefficients are analytical functions with respect
to the parameter $\lambda$
 and to the variable x is shown in the following lemmas.

\vspace{.2in}

\begin{func}

 Equation (59) defines a function $ \xi(z,\lambda)$ for all $z\in D
$, $\lambda \in G$
with the following properties: 1. $\xi(z,\lambda)$ is analytic in $
D\times G$
2.$ \xi(x_{1,2}(\lambda)),\lambda]=\mp2\sqrt{\tau_{0}}$
3.$ \xi^{'}(z,\lambda)\neq 0 $for all $ z\in D$
\end{func}

\vspace{.2in}

\begin{func}
 Coefficients $ a_{10}, a_{20} $ , defined by (60) , are analytic
functions
for all $ (z,\lambda) \in D\times G $.
\end{func}

\vspace{.2in}

The proof of lemmas 1,2 is based on Hartog's theorem.
On employing the integral equation method we obtain the asymptotic
representations
for the solutions of the equation (2).

\vspace{.2in}

\begin{func}
 Differential equation (2) for $z\in D$ and $ Re z \geq 0$ and $
\lambda\in D$
has four linear independent solutions, which have two derivatives and
are
given by

\begin{eqnarray}
U_{1,2}^{+}(z,\lambda,h)=A_{1,2}a_{1,2}^{0}U(\mp\frac{i\tau}{h},\exp(\
mp\frac{i\pi}{4})
\frac{\xi(x)}{\sqrt{h}})(1+O(\frac{1}{h^{2}}))\\
U_{3,4}^{+}(z,\lambda,h)=y_{3,4}(z,\lambda,h)(1+ \frac{1}{h^{2}})
\end{eqnarray}
where
\begin{equation}
A_{1,2}=\frac{2}{h^{1/4}}
\exp(\mp\frac{i\pi}{8}
     \pm\frac{i\tau}{h}
     \ln\frac{\exp(\mp i\pi/2)\tau}{he})
\end{equation}

For $ Re z\leq 0$ and $ z\in D$ , $ \lambda\in G$ differential
equation (2) has four
linear independent solutions:

\begin{eqnarray}
U_{1,2}^{-}(z,\lambda,h)=A_{2,1}a_{1,2}^{0}U(\pm\frac{i\tau}{h},-\exp(
\pm\frac{i\pi}{4})
\frac{\xi(x)}{\sqrt{h}})[1+O(\frac{1}{h^{2}})]\\
U_{3,4}^{-}(z,\lambda,h)=y_{3,4}(z,\lambda,h)[1+ \frac{1}{h^{2}}]
\end{eqnarray}
\end{func}

On employing the asymptotic formulas \cite{lo:rev} for the parabolic
cylinder functions
which are valid both for $ z\rightarrow\infty$ and $h\rightarrow 0$
we observe

\begin{eqnarray}
U_{i}^{+}(z)\sim y_{i}(z), z\rightarrow\infty,\\
U_{i}^{-}(z) \sim y_{i}(z), z\rightarrow -\infty
\end{eqnarray}

Connection formulas.

\vspace{.2in}

The solutions ${\bf
U}^{+}\equiv(U_{1}^{+},U_{2}^{+},U_{3}^{+},U_{4}^{+})^{t}$
and ${\bf U}^{-}\equiv(U_{1}^{-},U_{2}^{-},U_{3}^{-},U_{4}^{-})^{t}$
are related by
 a linear transformation of the form

\begin{eqnarray}
{\bf U}^{+}(z,\lambda,h)= A(\lambda,h){\bf U}^{-}(z,\lambda,h),
\end{eqnarray}

where $ A(\lambda,h)$ is $ 4\times 4$ matrix.
On differentiating equation (25) and setting $ \xi=0$ we get the
system of linear
 equations with respect to $ A_{ik}(i,k=1,...4)$.It suffices to
determine
 approximations for the coefficients $A_{ik}$.

\begin{equation}
A=\left(\matrix{
\frac{\sqrt{2\pi}\exp(\frac{\pi
p\tau}{2h}+\frac{i\tau}{h}\ln\frac{\tau}{he})}{\Gamma(
1/2+\frac{i\tau}{h})}&-i\exp(\frac{\pi\tau}{h})& 0 & 0 &\cr
i\exp(\frac{\pi\tau}{h})& \frac{\sqrt{2\pi}\exp[\frac{\pi
p\tau}{2h}-\frac{i\tau}{h}\ln\frac{\tau}{he}]}{\Gamma(
1/2-\frac{i\tau}{h})}& 0 & 0 &\cr
0 & 0 & 1 & 0 \cr
0 & 0 & 0 & 1 \cr
}\right)
\end{equation}

 {\bf  Acknowlegement}

The present work was supported by the Swedesh Institute.
The author is indebted to K.E.Thylwe for the attention to work.

\end{document}